# Assessment Issues in Mathematics: Design Science Approach for developing an Expert-System Based Solution

**Hussein Genemo**
College of Engineering and Science
Victoria University, Melbourne, Australia
Email: hussein.genemo@live.vu.edu.au

**Shah Jahan Miah**
College of Business, Victoria University
Melbourne, Australia
Email: shah.miah@vu.edu.au

## Abstract

Information and communication technology (ICT) tools are ineffective when assessing solutions of questions with more than one step in Mathematics. ICT tools assessing these types of questions are paralleled to solving complex problems. This conceptual paper describes a new combined approach to develop a rule based expert system (ES) prototype, implementing method marking concept (MMC) in assessing multi-step question (MSQ) solutions. Pragmatist views and mixed methods are jointly employed to conduct document analysis using undergraduate students' solutions to past examination papers and related findings from previous research. Design science research (DSR) paradigm and ES development phases will be used to design, construct and evaluate the artifact based on ES. The expected functioning ES artifact will enhance the authenticity of assessment, because evaluation of solution strategies in MSQ solution reveals the scope in the comprehension of the assessed domain knowledge. The artifact and the conceptual framework can be applied to a similar domain with modifications or as it is.

**Keywords**

Design Science Research, Expert Systems, Multi-step Question, ICT system, Method Marking Concept.

## 1   Introduction

Authentic assessment delivered using an appropriate assessment tool plays significant roles in enticing students towards attaining lifelong education (Bloxham & Boyd, 2007). Conventional ways of assessing Multi-step Questions' (MSQ) solution comprising steps to produce final answer is relatively easier, while using ICT tools to assess MSQ solution is similar in complexity to solving classified 'complex problem'. According to (Fischer, Greiff, & Funke, 2012) "trying to make a modern computer do what it is supposed to can turn out to be a complex problem" (p.36). These authors refer to well-structured problems that can be solved conventionally, without ICT application, becoming ill-structured problems when ICT solution is sought. This implies the fact that the complexity of a task depends on many factors.

Current ICT tools in mathematics assessment are limited to assessing only the final answer of MSQ's solution, but this type of assessment affects the authenticity of the content to be assessed, since the concern of evaluating MSQ's solution strategies applied is not addressed. This paper proposes conceptual solution to deal with this issue. Established theories related to complex problem solving paradigm, DSR methodology, pragmatist view and mixed methods methodology, (Waterman, 1986) and ES development phases are considered as the key determinants that guide to achieve the application of method marking concept (MMC) in assessing MSQ solutions. Researchers, as being one of research's instruments, their research skills also play vital roles in delivering quality outcome of research. The responsibility of selecting appropriate methods lies with researchers who "make choices regarding data sources, data construction, and analysis methods that best fit their research questions and to consider using multiple approaches and modes of inquiry" (Ercikan & Roth, 2006, p.55). In this research students' solutions to previous examinations and related documents are to be studied using pragmatist approach and mixed methods methodologies to elicit errors and solutions strategies from documents. DSR methodology in conjunction with ES development phases (Waterman, 1986) will use findings from the mixed methods investigation to build the ES artifact that will achieve MMC when assessing MSQs' solutions.

MMC is the notion that awarding learners' work is based on the solution strategies that are used in solving MSQ. DSR is a research paradigm that is used to build new artifacts that solve significant and applicable existing social and/or individual problems and at the same time add new acceptable knowledge to the science world (Hevner & Chatterjee, 2010, p. 15). The conceptual framework uses rule based ES to realise MMC and it is referred to as method marking concept rule based expert system (MMCRBES).

According to Hevner et al. (2004) any research studies that are designed for developing solution artefact, are classified as DSR. The research questions that from a DSR perspective are: What are the most common errors in algebraic MSQs solutions? What types of strategies are used in solving MSQ? How can MMC be applied to assess and award marks to steps in MSQ solutions, using an artifact based on ES technology?

This conceptual paper is structured as follows: section 2 describes educational assessment's roles in students' academic achievement. It also explains the relationships between processes of ICT tool assessing MSQ solutions and complex problem solving characteristics. The section also elaborates the research methodologies to be used. Section 3 deals with the methodologies to study documents required in the creation of MMCRBES knowledge and approaches to build and evaluate MMCRBES. Section 4 discusses the functionality of MMCRBES artifact prototype. Section 5 discusses the knowledge contribution of the artifact and the selected paradigm and methodologies to study the documents and section 6 summarises the conceptual paper and makes some suggestion for future study.

## 2    STUDY BACKGROUND

Authentic assessment focusing on measuring important skills considering "intellectual quality" and making improvement in students learning (Palm, 2008) plays important roles in influencing students' approach towards learning (Price, Carroll, O'Donovan, & Rust, 2011; Sangwin, 2012). Assessment is used to examine and report individuals' knowledge in a particular domain so that actions could be taken based on the outcome of the assessment. As the authenticity of assessment content is crucial, so too is the selection of an appropriate tool – which determines how and what to assess-. Assessment tools can control how and what to assess in solutions.

Assessing MSQ solution with ICT tool is similar to solving complex problems. ICT mathematics assessment tool has characteristics of complex problem solution. Yeo and Marqardt (2012) recommend problem solvers to accrue procedural ability and to possess enough information and strategies -not apparent in the problem to be solved- to face challenges of solving complex problem. According to Wüstenberg, et al. (2012) "complex problem solving is a process where individuals approach, manage, and address ongoing issues that are highly unpredictable" (p.263). It is hard for ICT tools to recognize in advance which solution strategies students use to solve MSQs. There is a lack of sufficient knowledge, available to the assessment tool, to evaluate MSQ solution. To solve these types of issues innovative skills, relevant knowledge and suitable approaches are required.

Based on the evaluation of relevant literatures and domain expert skills, MSQ assessment using ICT as an assessment tool reveals difficulties such as the absence of information that reveals, for example the number of steps required to produce final answers and also the order of steps used. . Furthermore there are issues such as correct algorithm producing incorrect step answer and incorrect answer usage in current step producing incorrect answer in subsequent steps regardless of the accuracy of algorithms applied. Many scenarios which are similar to these examples are prevalent when assessing MSQ solutions. These are some of the problems that current Mathematics ICT assessment tools have been facing.

Current ICT assessment tools used for mathematics are limited only to checking a final answer in MSQ solution evaluation (Beevers, Wild, McGuire, Fiddes, & Youngson, 1999;Livne, Livne, & Wight; 2015; Sangwin, 2012; Sangwin, Cazes, Lee, & Wong, 2010). The inability to assess students' whole work affects teaching and learning quality. It is important to ensure that students' understanding of a domain is fulfilled by exploring their knowledge through assessment that acts like a "bridge between teaching and learning" (Wiliam, 2005). To address the problem of assessing MSQs solutions, current ICT assessment tools redesigned these questions by breaking them into sub questions requiring only one stage to solve. However the concept of splitting complex questions raises problems with the assessment authenticity due to the facts that (1) the choice and use of algorithms to solve the question are not in themselves tested, because methods are shown to the students and (2) a particular step is forced upon the students to follow, which might be new or hard to use (Lawson, 2012).

This research aims to study scripts of first year undergraduate's previous solutions to algebraic examinations to extract errors and solution strategies and use the findings to develop MMCRBES artifact prototype that will implement MMC.As mentioned earlier our approach combines pragmatist paradigm and mixed methods methodology that will be used to study students' documents. DSR methodology and Waterman's ES development phases will guide the development of the ES artifact. The structures of the artifact design process and conceptual framework have been constructed based on the studies of related literatures and researchers' skills.

## 3    METHODOLOGY

In this section the methodology is described in two parts. First part is made up of two phases involving studying documents using qualitative and quantitative methods. Second part covers the process that will be used to design, build and evaluate MMCRBES artifact using the combination of DSR methodology and Waterman's ES development phases.

### 3.1    Phase 1: Qualitative Document Collection and Analysis

Documents will be gathered based on the criteria that they contain qualitative data/information to answer research questions that require the uncovering of common student errors and question solving strategies employed by students. The documents are previous students' solutions to examination papers that belong to two first year undergraduate mathematical foundations subjects. These documents are combined from five different semesters with different questions having similar academic content. Each document represents the work of a student who took only a single examination. From total of 300 papers, 100 papers will be used in our proposed qualitative study; equal number of papers, if possible, will be picked from each semester and subject to widen the representation of students works to be studied.

Researchers need "to develop a rigorous, systematic inquiry" (Creswell, Hanson, Clark Plano, & Morales, 2007, p. 236) to conduct research. This can be done by using qualitative or quantitative or mixed methods in a single research. More details about qualitative research are given in the subsequent paragraphs.

According to Guest, Namey and Mitchell (2013, p. 2) "there are about as many definitions of qualitative research as there are books on the subject. Some authors highlight the research purpose and focus". Creswell (2003) details the purpose of qualitative research as the research "to understand a particular social situation, event, role, group, or interaction. It is largely an investigative process where the researcher gradually makes sense of a social phenomenon by contrasting, comparing, replicating cataloguing and classifying the object of study" (Creswell, 2003, p.226). Creswell's study also suggested that "this entails immersion in the everyday life of the setting chosen for the study; the researcher enters the information's world and through ongoing interaction, seeks the informants' perspectives and meanings" (p. 226). Creswell's explanation of the function of qualitative research shows significant roles that qualitative research plays in revealing rich information about phenomenon. In the documents, the phenomenon that will be investigated is errors and solution strategies in the students' works.

Research question is considered as the driver of the whole research. According to Ellis & Levy (2008) "the nature of what is going wrong – the problem – very much sets the parameters for what can be done" (p. 22). Creswell et al. (2007) also state that research question "informs the approach or design used in qualitative research to collect and analyse the data" (p. 238). The most suitable research design is required to study the identified qualitative research question. According to Creswell et al. (20007) there are five prevalent qualitative research designs which highlight "the procedures involved in actually conducting qualitative studies"( p. 237). Creswell et al. (2007) refer to research design as "approaches to qualitative research that encompass formulating research questions and procedures for collecting, analysing, and reporting findings" (p. 237). From the five qualitative research designs, the phenomenology design seems appropriate to collect and analyse documents. However the inductive thematic analysis suggested by Guest, Namey & Mitchell (2013) seems more proper for its procedure is "consists of reading through textual data, identifying themes in the data, coding those themes, and then interpreting the structure and content of the themes"(p. 13).

Qualitative research has "an inductive and flexible nature" (Guest et al. 2013, p. 4). And the flexibility aspects offers prospect to researchers to alter "the sampling procedures during the data collection process based on incoming data" (Guest et al. 2013, p. 4). In qualitative research, generally, the research questions are open-ended type to enable researchers to probe the problem for further information; this process of probbing for more information leads to producing valid research data (Guest et al. 2013, p. 4).

The study of examinations documents will be approached with this type of open-ended relevant questions in mind to thoroughly investigate the papers iteratively. This qualitative research objectivities of identifying and exploring errors and solution strategies, describing, and explaining them will assist in accumulating the knowledge that is needed by the expert system to construct MMCRBES. These objectives can be contrasted with the purposes of qualitative research identified, above, by Creswell (2003).

However there are weaknesses that are pertained to qualitative study. Time, resources and sampling issues are sensitive, because for example "proper analysis of text is time consuming"(Guest et al. 2013, p. 25) and "the ability to claim a representative sample is often diminished, and statistical generalization is impossible" (Guest et al. 2013, p. 25). The other concern is that qualitative research is "not an ideal choice for reliably comparing groups" (Guest et al. 2013, p. 25). That is why when it comes to finding common errors and solution strategies, the qualitative data will be transferred to quantitative data and interpreted quantitatively.

### 3.1.1 Documents Data Analysis process

The key objective of this qualitative method phase is to examine documents. Thematic analysis is chosen to analyse the papers. This type of analysis enables "the researcher to compare and contrast the emergent themes throughout the process" (Rasmussen, Cochrane & Henderson, 2012); the emergent themes will be students' errors in the solutions and methods of solving questions. The validity of the qualitative method, to collect and analyse documents, will be enhanced by "documenting the relationship between data sources and a study's research questions, the development of themes and categories, and the triangulation of findings"(Anfara, Brown & Mangione, 2002) using tabular system. The other option that this current research might apply is cross-referencing (Anfara, Brown & Mangione, 2002) questions- in the documents - to be selected to the curriculum of the domain subject to see if the questions that their solutions are to be studied covers the knowledge in the domain.

### 3.1.2 Phase 2: Quantitative Document Collection Analysis

About 200 undergraduate students' previous examination solution papers- expanding over two years- will be accessed and analysed using quantitative method and by exploiting the findings resulted from the qualitative method. Descriptive or summary statistics will be used to determine the similarity or difference between years, to isolate causes and effects, and to eliminate outliers. In this research one of the main interests is to reveal the number of time particular errors and solutions strategies repeated in the documents to be studied. The descriptive or summary statistics show promises to discover that common features emerged across such documents. Since mixed methods are used in the research, whatever emerged as themes in quantitative research is highlighted upon in the qualitative part of the research and investigated further.

## 3.2 Solution design

As mentioned earlier a combined approach of DSR and Waterman's (1986) methodologies will be used for the ES prototype development. The design science focuses on studies of how things might be instead of how things are (Chatterjee, 2010; Hevner & Chatterjee, 2010; Neuhauser et al., 2013). It is viewed as the research paradigm that contributes to the research that produces an artefact that is useful in solving appropriate human problems and at the same time adds new knowledge to the body of scientific evidence (Chatterjee, 2010; Hevner, et al., 2004; Neuhauser et al., 2013).

In DSR methodology, behavioural science (BS) and design science disciplines are regarded complementary (Hevner, March, Park & Ram, 2004). Figure 1 reflects this reality in which BS activities are represented by continuous line rectangles. According to Hevner (2007, p. 88), any genuine design science research should contain relevance, design and rigor cycles to deliver purposeful artifacts. The relevance cycle provides research questions and criteria to test artifacts while the design cycle is where the building and evaluation of artifacts take place. The rigor cycle "connects the design science activities with the knowledge base of scientific foundations, experience, and expertise that informs the research project" (p. 89).

BS is required to study the phenomena in the environment where the design science research questions and evaluation criteria are sourced. Pereira et al. (2013, p. 152) view that phenomena and environment are inseparable entities. On the other hand, "design science provides substantive tests of the claims of natural science research" (March & Smith, 1995, p. 255). These practical tests are conducted through exploration of artifacts. Therefore, "design science complements behavioral science in that the former focuses on the design and creation of artifacts; and the latter, on the observation and evaluation of the characteristics of these artifacts and how they relate and interact with users and organizations, in general"(Shaw & Piramuthu, 2008, p. 715). In current research behavioral science approach will be applied to investigate research problems to form data for problems' solutions and evaluation criteria.

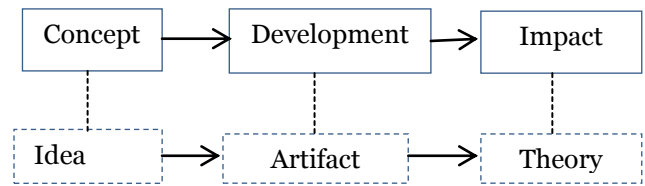

*Figure 1. DSR and BS complementation and interaction*

One of the data sources, in figure 3, is documents analysis findings that will be used to build constructs/concepts which "form the vocabulary of a domain" (March & Smith, 1995, p. 256). According to Hevner et al. (2004) "constructs provide the vocabulary and symbols used to define problems and solutions. They have a significant impact on the way in which tasks and problems are conceived. They enable construction of models or representations of the problem domain" (Hevner et al. 2004, p. 83).

The developed constructs will be used to explain and visualise the nature of the phenomena data - errors and solution strategies- and the way the data is formed. The attributes in the data determine the structure of the classification of the data in categories. At classification phase, categories from the initial observation and description of the phenomena are formed (Chatterjee, 2010, p. 35). The categories structure -data format- created from the data in the phenomena shows the relationship between data in the phenomena and the intended outputs which are algorithms that will be used to build production rules for MMCRBES artifact that is based on ES.

ES is an existing artifact and in this research it will be used in the development of MMCRBES artifact that will assess algorithms used to produce MSQ answers. It is hard to know in advance methods used to solve these questions; the solution could be incomplete. It is difficult to apply conventional programming technique to assess the whole work related to the solutions of this type of questions. Using ES's reasoning feature might provide the solution to this predicament (Giarratano & Riley, 2005, p. 23).

### 3.2.1 ES as Solution

ES is a specific form of artificial intelligent software that is specialized to solve complex particular domain problems or assist in making decision in a particular domain emulating human experts (Giarratano & Riley, 2005; Sasikumar, Ramani, Raman, Anjaneyulu, & Chandrasekar, 2007; Turban, Aronson, & Liang, 2005). Miah, Kerr & Gammack (2009) describe three directions of ES as advisory systems which focus on replacing human expertise; diagnostic systems which focus on symptom/clue based knowledge stored; and planning and management support systems which focus on management support with appropriate knowledge. Implies form Miah et al. (2009), all types of ESs require knowledge to operate. There are three main categories of knowledge. These are: declarative, procedural and metaknowledge (Turban, et al., 2005, p. 582).

Declarative knowledge is identified as the type of knowledge that is fact and truth and it is represented in the form of factual statement that someone can understand and rationalise it (Anderson, 1988; Giarratano & Riley, 2005; Stalnaker, 2012). Anderson (1988) claims that procedural knowledge is developed from declarative knowledge. The author also states "that knowledge must start in a declarative form before becoming peroceduralised" (p.41). Stalnaker (2012) argues the differences between procedural and declarative knowledge arise from the way they are used and this difference has root in computer science. The author explains the reason "was to distinguish knowledge that is explicitly encoded as data and knowledge that is realized in the program or in the structure of the system" (p. 757). Declarative knowledge is required in all stages of expert systems cycle, for example domain experts share facts using declarative knowledge during knowledge acquisition phase.

Procedural knowledge is the knowledge that requires information about doing things (Jaques et al., 2013). "Metaknowledge is the knowledge about knowledge" (Turban et al., 2005, p.583) and having knowledge about the procedure of knowledge based systems is regarded as meta-knowledge (Turban et al., 2005).

The MMCRBES artifact development needs all the three types of knowledge mentioned above. The findings from the qualitative and quantitative research are expressed in declarative form. Furthermore this knowledge will be formatted and stored as data in the knowledge base in declarative form. MMCRBES artifact also requires procedural knowledge that can be formalized into computer algorithm (Jaques et al., 2013). The MMCRBES artifact needs both procedural and metaknowledge knowledge. It requires procedural knowledge, because mathematics content is procedural and it is also easy to formalize procedural knowledge into computer algorithms (Jaques et al., 2013). The metaknowledge knowledge assists in understanding and developing the MMCRBES artifact. Rule based ES implements procedural rule (Angeli, 2010) and this is why MMCRBES artifact prototype will be created using the rule-based ES.

### 3.2.2　MMCRBES Development Cycle

The MMCRBES will be created using Hevner et al. (2004) DSR's seven guidelines and Waterman's (1986) ES development phases. Waterman's ES development phases are specific to the development of ES. On the other hand Hevner's guidelines are general and applicable to any information system (IS) research. Waterman's development phases are: identification, conceptualization, formalization, implementation and testing. Here the link that will be made between Hevner's guidelines and Waterman's ES development phases is to show how the two approaches are utilized in the development of the MMCRBES artifact.

In the identification phase- conforming to Hevner's et al. (2004) "design as artifact and problem relevance" guideline- the research requirements can be identified through the research questions' phenomena studies and literatures review which identify related knowledge and issues that require new investigation. The conceptualization phase conforms to Hevner's et al. (2004) "research rigor and design as a search process" guidelines. In this phase elements in the document analysis findings are searched and relationships between them are established; interactions between elements and interactions' controls also determined in this phase.

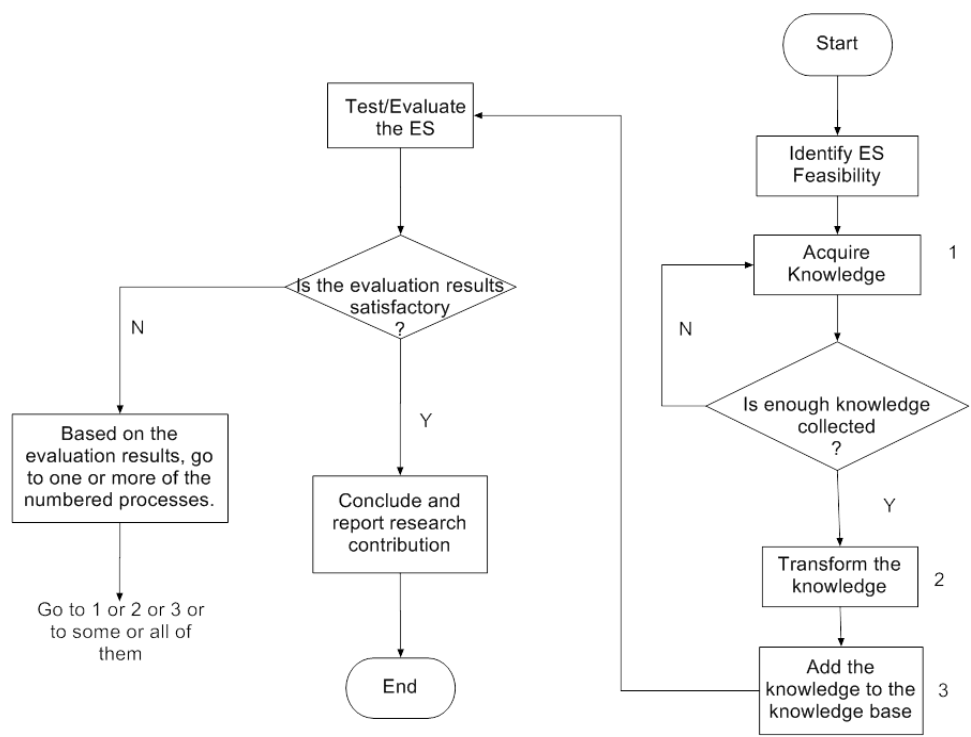

*Figure 2: MMCRBES Development Phases*

The formalization phase - conforms to Hevner's et al. (2004) "design as a search process" guideline- is the stage where the knowledge that is obtained in the conceptualisation phase is transformed into the format that is useable by ES. At the implementation phase - conforming to Hevner's et al. (2004) "design as a search process" guideline-, the formatted knowledge is stored in the ES for operation. The testing phase - conforms to Hevner's et al. (2004) "design evaluation" guideline– performs the verification of elements existence, validity of relationships and interactions and the artifact's performance and utility. Please refer to table 1 for the details of the processes in the flowchart of figure 2. The flowchart sequences and activities are based on the Waterman's ES development phases and Hevner's et al. (2004) DSR guidelines.

| | |
|---|---|
| **Identfiy ES Feasibility** | Identifying the requirements for the possible devlopment of ES. |
| **Acquire Knowledge  1** | Continue Extracting knowledge from document analysis findings (errors and solution algorithms and classifying them into groups) till satisfied that engough knowledge is acquired. |
| **Transform the Knowledge   2** | Transform the acquired knowledge and do this rigorously till sufficient knowledge is transformed into the format that is recognized by ES. |
| **Add the Knowledge to the Knowledge Base  3** | Put the knowledge into the knowledge base and run the ES prototype. This is the implementation phase of the ES. |
| **Test/Evaluate the ES** | Test the system for the processes in the boxes numbered 1-3. Also validate the performance and utility of the ES. |
| **Based on the evaluation results, go to one or more of the numbered processes. 1, 2, 3** | If the evaluation feedback requires the modification, go to the processes 1, or 2 or 3 or combination of them to satisfy the requirements in the feedback returned by the evaluation/test phase. |
| **Conclude and report the research contribution** | This is shows the end of successful artifact development and as the result, communicate the research contribution to knowledge base. |

*Table 1. Details of processes in the MMCRBES Development Phases flowchart*

# 4   PROPOSED ES (conceptual framework)

The MMCRBES architecture that is shown in figure 3 includes knowledge base, knowledge (data) source, blackboard, inference engine, and user interface and users components. The information- facts and rules- in the knowledge base that is required for the MMCRBES artifact operation is constructed from the knowledge source. The blackboard, also called working memory,  module in rule-based expert system is a medium acting as a database and its purpose is to store information about specific case of current problem that ES is focusing on (Sasikumar et al., 2007; Turban et al., 2005; Waterman, 1986). The inference engine takes input from users and attempt to resolve problems using the control in the rule base and the knowledge in the working memory (Sasikumar et al., 2007). The purpose of user interface is to allow users interaction during MMCREB artifact's development and functioning.

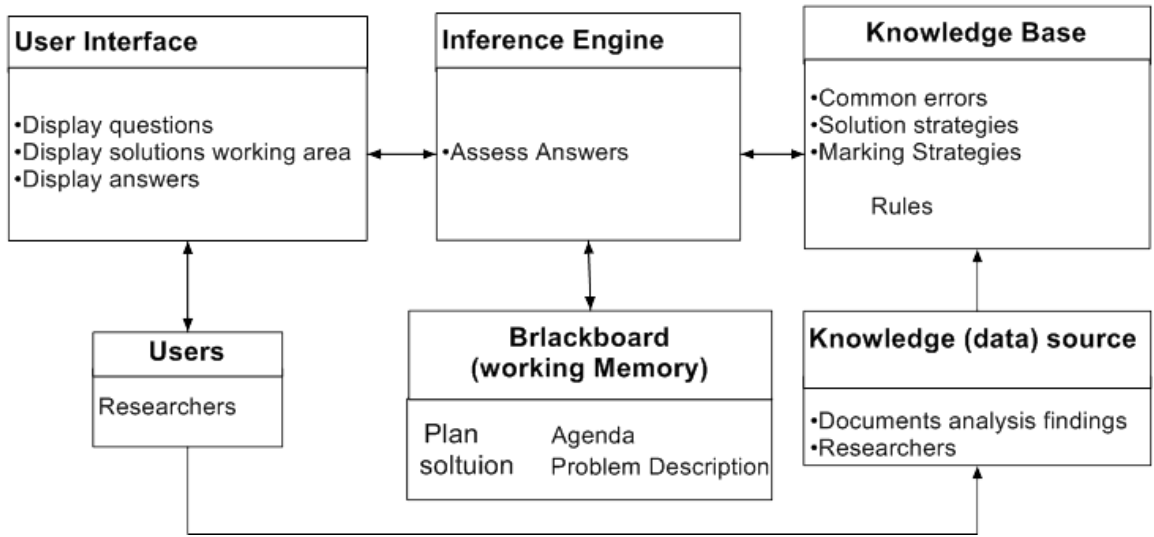

Figure 3. Overall architecture of the proposed solution artefact

## 4.1 Assessing MSQ Questions' Solutions

The operational process- shown in flowchart of figure 4 of MMCRBES artifact to assess MSQ solution is explained next. Here the inference engine using the available knowledge determines if there is knowledge that can be used for the step solution submitted by a user. If there is no solution, record the reason for the absence of the solution and check if all the steps that are required to answer the question have fully been used. If the knowledge to assess current step is available, assess the solution in the step, assign mark to it and give choice to users to continue to the next step or to end answering the question. If all the steps are completed or users requested to end the assessment process- before answering the question fully- add steps' marks, display the total mark and end the assessment process.

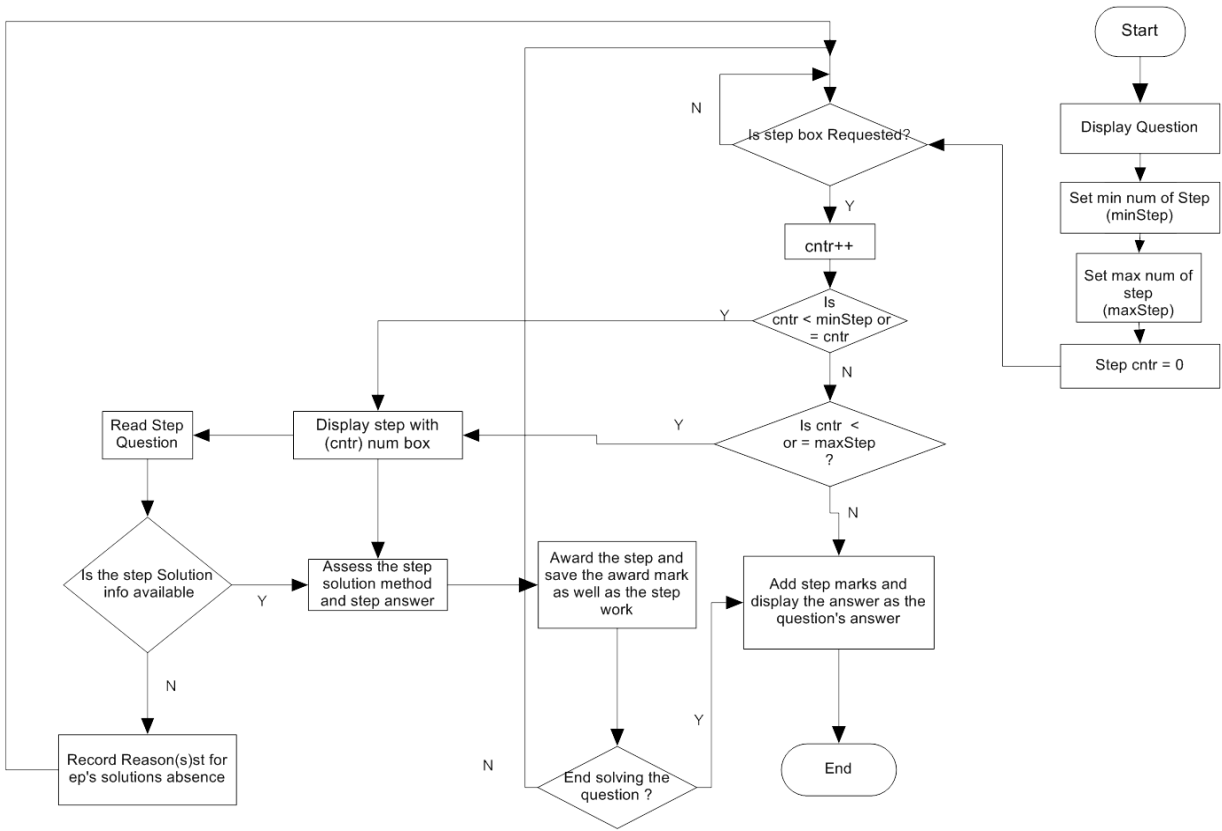

Figure 4. Working principle of the proposed solution artefact

The MMCRBES artifact function is based on providing questions, answer and feedback and working areas for users. Users select the number of steps that current question require to produce the final answer and go through each step to work on the step's answer. The final step would be the answer the answer of MSQ. At any time users can end working on current step's solution and submit the whole answer's to end current question's answering process or session.

The system displays each step's answers based on the number of steps answered. The ES artifact assesses and awards answers to steps' in questions based on the criteria set by the assessors who have the privilege to prepare question and assessment criteria and store them in the system. At the end of the session or upon the request of users' current question's answers with feedback is displayed. The system allows developers to add features to the MMCRBES artifact functionality such as modifying or adding knowledge to knowledge base or setting new rules. If the artifact is not able to assess user's submitted solution, the artifact saves the submitted solution giving opportunity to developers or/and assessors to review and take actions that enhance the artifact functionality.

## 5  DISCUSSION AND EVALUATION

This research in progress is to develop MMCRBES artifact that assesses MSQ solutions and also to add new knowledge to the body of the target literature. Although at this level of developing an initial conceptual knowledge, the paper defined the MMCRBES artifact which is competent to implement MMC and make valuable contribution to the domain community. Assessing learners' full work reveals major advantages to all stakeholders who are learners, educators, policy makers and any allied bodies related to education of the domain and associated matters. An Assessment outcome can influence teaching and learning methods and contents. The same concept can be applied in similar domains, such as physics, chemistry and any other discipline where complex problems solutions are required. There are many factors that might affect the functional characteristics of MMCRBES to be built; these are the data source selection criteria, methods and types of knowledge extracted and the extent of design and evaluation involved. Researchers will make all possible endeavours to minimise limitations associated to the functionality of MMCRBES artifact. In this case, methodological guidelines such as Waterman's ES development phases and Hevner's DSR guidelines are just recommendations. Every possible and applicable method will be used to design and create MMCRBES artifact that is purposeful.

Pragmatism is the most appropriate paradigm to conduct this research. It is not restricted to any particular philosophy giving freedom to researchers to select "the methods, techniques, and procedures of research that best meet their needs and purposes"(Creswell, 2003, p. 12) and it enables them "to look beyond the "what" and "how" to research based on its intended consequences-where they want to go with it" (Creswell, 2003). In pragmatism the focus is on the problem and the solution rather than the methods used to investigate a research question(s). The "concern is with applications "what works" – and solutions to problems" (Creswell, 2003, p. 11). Pragmatism paradigm is associated with mixed methods methodology. Pragmatism paradigm is associated with mixed methods methodology; and according to Johnson and Onwuegbuzie, (2004) the bottom line is that approaches are to be mixed in such a way that may offer the best opportunities for answering our research questions and that is what MMCRBES artifact development needs.

In mixed methods approach, the qualitative and quantitative strategies are used to collect and analyse data using both qualitative and quantitative methods in a single study (Johnson and Onwuegbuzie, 2004, Creswell, 2003). In this research the priority will be given to the data in the quantitative approach. The sequential exploratory technique will be used to study the phenomenon. The purpose is to expand the findings from the qualitative phase output in the second phase by applying quantitative data collection and analysis methods. During the interpretation phase the two findings will be integrated to produce the data that will be used in building and evaluating MMCRBES artifact. This type of mixed method approach is applied "to generalize qualitative findings to different samples" (Creswell, 2003, p. 215). In this part of the research the main purpose is to extract common errors made and solutions strategies which are implemented in MSQs solutions. Researchers' proper usage of pragmatism paradigm and associated mixed methods and design science research paradigm hopefully expected to deliver MMCRBES artifact that will be effective in implementing MMC.

# 6　CONCLUSION

Educational authentic assessment using proper assessment tools influences learning and instruction process. However, majority of the assessment tools, used for assessment, produce similar results. ICT tools for assessing MSQ solutions in this instance failed to appropriately evaluate learners' solution strategies that reveal the real comprehension of the domain under evaluation. The existing ICT tools resolve the problem of MSQ solution assessment by simplifying questions. This action eliminates solution strategy assessment leaving the assessment authenticity doubtful. This research attempts to improve this issue through the implementation of MMC that facilitates the assessment of student's solutions strategies. be able to assess solution methods applied Pragmatist view together with mixed methods methodologies are expected to assist in collecting and analysing students' previous examinations' solutions papers and related previous documents.

The outcome of document analysis will be studied further and transformed to the knowledge that will be used in the design, creation and evaluation of ES. The study of functioning conceptual framework shows benefits of the ES artifact in serving the domain community that will use the ES artifact. It is expected as any artifact nature, limitations may be persistent due to the nature of complex problem's unpredictability and shortfalls of ES reasoning. In addition, ES solutions as assessing tool may not be fully functioned as human assessors do. Future research can add aspects such as collecting all the errors learners make during MSQ solutions, study them and subsequently create protocols and build errors' repository. The other possibilities are incorporating techniques of complex problem solutions and cognitive knowledge in studying, designing and creating ES artifact. Another opportunity is that the MMCRBES artifact (design, evaluation, prototype) aspects itself being the subject of new research in the other disciplinary areas.